\documentclass
[aps,prb,twocolumn,showpacs,amsmath,superscriptaddress,amssymb]{revtex4}%
\usepackage{graphicx}
\usepackage{dcolumn}
\usepackage{bm}
\usepackage{amsmath}
\usepackage{amsfonts}
\usepackage{amssymb}
\usepackage{color}%
\providecommand{\U}[1]{\protect\rule{.1in}{.1in}}

\providecommand{\U}[1]{\protect\rule{.1in}{.1in}}
\begin{document}
\title{Manipulation of Ferromagnets by the Spin-Selective Optical Stark Effect}
\author{Alireza Qaiumzadeh}
\affiliation{Department of Physics, Norwegian University of Science and Technology, NO-7491
Trondheim, Norway}
\author{Gerrit E. W. Bauer}
\affiliation{Institute for Materials Research, Tohoku University,
Sendai 980-8577, Japan}
\affiliation{WPI Advanced Institute for
Materials Research, Tohoku University, Sendai 980-8577, Japan}
\affiliation{Delft University of Technology, Kavli Institute of
NanoScience, 2628 CJ Delft, The Netherlands}
\author{Arne Brataas}
\affiliation{Department of Physics, Norwegian University of Science and Technology, NO-7491
Trondheim, Norway}

\begin{abstract}
We investigate the non-resonant all-optical switching  of
magnetization. We treat the inverse Faraday effect (IFE)
theoretically in terms of the spin-selective optical Stark effect
for linearly or circularly polarized light. In the dilute magnetic
semiconductors (Ga,Mn)As, strong laser pulses below the band gap
induce effective magnetic fields of several teslas in a direction
which depends on the magnetization direction as well as the light
polarization and direction. Our theory demonstrates that the
polarized light catalyzes the angular momentum transfer between the
lattice and the magnetization.

\end{abstract}

\pacs{75.78.-n, 78.20.Bh, 71.70.Ej, 75.40.Gb}
\date{\today}
\maketitle

\section{Introduction}
An essential challenge in magnetoelectronics is finding new methods
to manipulate the magnetization and increase the switching speed for
realizing faster and higher-density data storage and information
processing. Traditionally, the magnetization is switched by applying
nearly collinear external magnetic fields for longer than
$100~\mathrm{ps}$ \cite{siegmann}. Current-induced spin transfer
torque switching is not faster \cite{stt}. So-called precessional
switching can be achieved by a magnetic field pulse perpendicular to
the magnetization \cite{siegmann, ultimate speed}, leading to
magnetization reversal as fast as, but not faster than, a few
$\mathrm{ps}$ \cite{ultimate speed}. In this paper, we explain how
off-resonant optical pulses generate strong effective magnetic
fields that can lead to ultrafast magnetic reversal.

Recently, it has been demonstrated that a $40~\mathrm{fs}$
circularly polarized pulse can reverse the magnetization of the
metallic ferrimagnet GdFeCo with perpendicular magnetic anisotropy
\cite{kimel1}. A magnetic write event as short as $30~\mathrm{ps}$
by using $100~\mathrm{fs}$ circularly polarized light pulses has
been reported \cite{kimel2}, \textit{i.e.} a potential data storage
rate of about $10$\textbf{ }$\mathrm{Tbit/s}$. Simulations suggest
that the magnetization reversal is not realized via precession, but
is caused by a linear process in which the magnitude of the
magnetization passes through zero, in the presence of magnetic
fields of $\sim$ 20 T \cite{kimel1,kimel2}. The magnetization
dynamics can also be triggered by light that is linearly polarized
in a direction non-collinear to the crystal axis \cite{icme2,icme3}.
For a comprehensive review on ultrafast all-optical magnetization
dynamics, see Ref. \cite{rmp_kimel}. Phenomenologically, these
experiments are attributed to two reactive effects: (i) the inverse
Faraday effect (IFE) \cite{pershan,shen} - the ability of the
electric field component of circularly polarized light
$\mathbf{E}(\omega)$ to induce a
static magnetization $\mathbf{M}_{\mathrm{IFE}}(0)\propto\mathbf{E}%
(\omega)\times\mathbf{E}^{\ast}(\omega)$ - and (ii) the inverse Cotton-Mouton
effect (ICME) \cite{shen} - the magnetization induced by polarized light in
the presence of an external magnetic field $\mathbf{{B}}_{\mathrm{ext}}$,
$\mathbf{M}_{\mathrm{ICME}}(0)\propto|\mathbf{E}(\omega)|^{2}\mathbf{{B}%
}_{\mathrm{ext}}$.

The microscopic origin of the large magnetic field induced by light
and the induced magnetization dynamics are still a subject of debate
\cite{rmp_kimel,tailoring light,modeling}. Several theoretical
mechanisms have been proposed, such as the optical Barnett effect or
the inverse Einstein-de Haas effect \cite{barnett}, light-induced
circular currents in the collisionless limit
\cite{kubo_plasma,Fateme}, the impulsive stimulated Raman-like
process \cite{rmp_kimel}, and photonic angular momentum transfer via
deflection of the scattered photons \cite{Woodford}. A dissipative
IFE under THz irradiation has been computed for dirty metals with
extrinsic spin-orbit interaction \cite{edelstein,tatara}. Also
experimentally the situation is not clear. Recent experiments on the
ferrimagnetic metallic alloy GdFeCo found optical magnetization
switching without light polarization in a certain range of light
intensities and sample temperatures, casting doubt on the ubiquity
of the IFE \cite{deJong}. 
Also Such a behavior was not reported for TbCo films, however
\cite{Mangin}.

In this paper, we predict huge effective magnetic fields induced by
the below band gap polarized light through the spin-selective ac
(optical) Stark effect \cite{stark_ife}, \textit{i.e}. the shift of
electronic energy levels connected through finite optical matrix
elements \cite{shen}. In perturbation theory, this process is
closely related to Raman scattering. Since the electronic structure
of amorphous alloys is complicated and experiments concerning the
role of the IFE are inconclusive, we focus here on GaMnAs as a
generic model system, since it can be grown with perpendicular
anisotropy and its electronic structure is well known. Dilute
magnetic semiconductors are interesting spintronic materials by
themselves \cite{awschalom}. Although their Curie temperature at
present is below the room temperature, studying these materials can
improve our understanding of novel physical phenomena that are also
present in other magnets \cite{munekata,jungwirth,phelps}.
Photo-injected carriers induced by linearly polarized light with
frequency slightly above the $\Gamma$ or L band edges have been
shown to induce magnetization dynamics in GaMnAs
\cite{munekata,perakis}. In contrast, we focus here on excitation
with frequencies below the fundamental band gap, which is
dissipationless, since no free carrier are excited. Our approach is
quite general and can be applied to arbitrary electronic structures
and computed from first principles.

In one scenario \cite{kimel1,kimel2}, extrinsic processes due to the
high-intensity laser pulse heat up the ferromagnet so that it
becomes paramagnetic while the circularly polarized light generates
the spin-selective optical Stark effect or effective magnetic field
$B_{\mathrm{IFE}}$ that triggers linear reversal, \textit{i.e}. the
modulus of the magnetization passes through zero during switching.
The maximum achievable field can be huge, \textit{e.g.} for GaAs
parameters $B_{\mathrm{IFE}}=\hslash \Omega/\left(
g_{s}^{\ast}\mu_{B}^{\ast}\right)  \sim100$ T, where $\Omega$
is the Rabi frequency for a light intensity of 5 GWcm$^{-2}$, and $g_{s}%
^{\ast}\mu_{B}^{\ast}/\hslash\sim500$ GHz/T is the effective
gyromagnetic ratio. In practice, the light frequency should be
sufficiently below the band gap to reduce heating that destroys the
sample at high intensities. We therefore formulate the IFE here in
second order perturbation theory and compute the resulting
expressions for the GaAs band structure and wave functions. We find
that for light frequencies safely below the energy gap, the
effective field amounts to several teslas in GaMnAs, which suffices
to nucleate a ferromagnetic state during the cooling phase \cite{kimel2,nowak}%
. In a second scenario, we assume that the material remains
ferromagnetic under the laser excitation, possibly with a reduced
magnetization. In this case, both linearly and circularly polarized
light can trigger both precessional and linear switching mechanisms.
In both scenarios, the required angular momentum is not supplied by
the photons, but by the lattice via the spin-orbit interaction. We
show that, in general, the light-induced effective field has three
components, depending on the light polarization and initial
magnetization direction, \textit{viz}. $\mathbf{{B}}_{\mathrm{light}%
}=B_{\mathrm{IFE}}\hat{\mathbf{{q}}}+B_{\mathrm{ICME}}~\hat{\mathbf{{M}}}%
_{0}+B_{\bot}\hat{\mathbf{{q}}}\times\hat{\mathbf{{M}}}_{0}$. The
sign of the IFE effective field $B_{\mathrm{IFE}}$ depends on the
helicity and points along the light propagation direction
$\hat{\mathbf{{q}}}$. $B_{\mathrm{ICME}}$, the magnetic field
associated with the ICME, is directed along the magnetization vector
$\hat{\mathbf{{M}}}_{0}$, while a field with strength $B_{\bot}$ is
perpendicular to both.

This paper is organized as follows. In Sec. II we present the Kane
band model we have used to describe light-matter interaction and
introduce the effective hamiltonian within the second order time
dependent perturbation theory. In Sec. III we present and discuss
our main analytical results for the light-induced magnetic field in
dilute magnetic semiconductors. Finally, in Sec. IV we present a
summary of our main conclusions.

\section{Model Hamiltonian and second-order time-dependent perturbation theory}
In the Coulomb gauge, the 8-band Kane model Hamiltonian for a
zinkblende
semiconductor at the $\Gamma$ point reads $\mathcal{H}=\mathcal{H}%
_{0}+\mathcal{H}_{\mathrm{p-d}}+\mathcal{H}_{\mathrm{int}}$, where
\begin{align}
\mathcal{H}_{0}  &  =\frac{{\mathbf{p}}^{2}}{2m}+\frac{\hbar}{4m^{2}c^{2}%
}\mathbf{p}\cdot\left(  \boldsymbol{s}\times\boldsymbol{\triangledown}%
V_{p}\right)  +V_{p},\nonumber\\
\mathcal{H}_{\mathrm{sp-d}}  &  =-J{\mathbf{\hat{M}}}_0\cdot\boldsymbol{s}%
,\nonumber\\
\mathcal{H}_{\mathrm{int}}  &  \simeq\frac{e}{mc}\mathbf{A}\cdot
\mathbf{p}+\frac{e^{2}}{2mc^{2}}{\mathbf{A}}^{2}. \label{hamiltonian}%
\end{align}
Here $\boldsymbol{p}$, $e$, and $m$ are the momentum operator,
electron charge, and electron mass, respectively, $\hbar$ is the
reduced Planck constant, $c$ is the light velocity, $\boldsymbol{s}$
is vector of $2\times2$ Pauli matrices, $\mathbf{A}$ is the vector
potential of the monochromatic light field, $V_{p}$ is the periodic
lattice potential, and $\mathcal{H}_{\mathrm{sp-d}}$ is the sp-d
mean-field exchange interaction between the magnetization direction
of the localized d-spins $\hat{\mathbf{{M}}}_0$, and the itinerant
s- or p-spins, controlled by the exchange potential $J$. The
$\mathbf{A}\cdot\mathbf{{p}}$ interaction term describes the
annihilation of a photon and the creation of an electron-hole pair
and \textit{vice versa}, while ${\mathbf{A}}^{2}$ represents a
photon scattering processes. In perturbation theory, two-photon
transitions can be induced by either $\mathbf{A}\cdot\mathbf{{p}}$
to second order or ${\mathbf{A}}^{2}$ to first order in the
interaction Hamiltonian. To leading order in the light-matter
interaction, ${\mathbf{A}}^{2}$ does not induce
spin reversal and will therefore be disregarded in the following. $\mathbf{A}%
\cdot\mathbf{{p}}$ induces only two-photon virtual interband
transitions, since the light frequency is below the band gap.
Intra-band transitions are disregarded because they are
impurity-mediated and weak. Direct-band-gap semiconductors can be
treated in the effective mass approximation and projected on the
well-established 8-band Kane model for $\mathcal{H}$ including the
conduction ($|cb^{\pm}\rangle$), the heavy-hole ($|hh^{\pm
}\rangle$), the light-hole ($|lh^{\pm}\rangle$), and the spin-orbit
split-off ($|so^{\pm}\rangle$) bands \cite{kane}. In the following,
we will disregard the band dispersion, a common approximation in
theories of Raman scattering \cite{pinczuk} that is allowed for low
doping levels and/or large detuning.

The electric field component of monochromatic light with frequency $\omega
_{0}$ and wave-vector ${\mathbf{q}}\Vert{\mathbf{\hat{z}}}$ is $\mathbf{{E}%
}(t)=\mathbf{{\hat{e}}}E_{0}e^{-i(\omega_{0}t-{\mathbf{q}}.{\mathbf{r}}%
)}+\mathrm{c.c.}$, for light propagating along the $z$-direction
with polarization
$\hat{\mathbf{{e}}}=e_{x}\hat{\mathbf{{x}}}+e_{y}\hat
{\mathbf{{y}}}$. For circular polarization
$e_{x}=1/\sqrt{2},e_{y}=\lambda i/\sqrt{2}$, where $\lambda=\pm1$,
and for linear polarization with angle $\alpha$ relative to the
$x$-axis, $e_{x}=\cos\alpha$ and $e_{y}=\sin\alpha$. When the pulse
duration $T_{p}$ is sufficiently longer than $\left(  E_{g}/\hbar
-\omega_{0}\right)  ^{-1}$, where $E_{g}$ is the energy gap,
transient effects
can be disregarded \cite{blugel}. For a laser pulse width of 40-100 $%
\operatorname{fs}%
$ with a frequency that is not too close to the resonance, the above criterion
is satisfied by $\hbar/E_{g}\approx0.5%
\operatorname{fs}%
$ for our material.

The matrix elements of the Hamiltonian in second order perturbation
for the $\mathbf{A}\cdot\mathbf{{p}}$ interaction term read
\cite{pershan}
\begin{align}
\langle m|\mathcal{H}|k\rangle &  =\frac{e^{2}E_{0}^{2}}{m^{2}\omega_{0}%
^{2}c^{2}}\sum_{l}\left[  \frac{\langle m|p_{\beta}e_{\beta}^{\ast}%
|l\rangle\langle l|p_{\gamma}e_{\gamma}|k\rangle}{\hbar\omega_{0}%
+(\epsilon_{k}-\epsilon_{l})}\right. \nonumber\\
&  \left.  -\frac{\langle m|p_{\gamma}e_{\gamma}|l\rangle\langle l|p_{\beta
}e_{\beta}^{\ast}|k\rangle}{\hbar\omega_{0}-(\epsilon_{m}-\epsilon_{l}%
)}\right]  ,
\end{align}
where $\beta(\gamma)=x,y,z$. $|m\rangle$, $|k\rangle$, and $|l\rangle$ are the
initial, final, and intermediate states including the spin and momentum
quantum numbers, with the energies $\epsilon_{m}$, $\epsilon_{k}$, and
$\epsilon_{l}$ respectively.

\section{light-induced Effective magnetic field in $\textrm{GaAs}$ and $\textrm{GaMnAs}$}
Unlike an external magnetic field, conduction and valence bands
experience different light-induced effective fields \cite{reid}. For
conduction band (valence bands)
$\mathbf{B}_{\mathrm{light}}=2\mathrm{Tr}\left[
\mathbf{s}\mathcal{H}\right]  /\left(  \mu_{B}^{\ast}g_{s}^{\ast}%
\mathrm{Tr\,}[\mathbf{s}{^{2}}]/3\right)  $, where $\mathbf{s}$ is
the vector of $2\times2$ Pauli spin matrices for $1/2$-spins in the
conduction band (the $4\times4$ spin matrices for $3/2$-spins in the
valence band), $\mathrm{Tr}$ is the trace over electron states (hole
states), $\mu_{B}^{\ast}$ is effective Bohr magneton of electron
(hole), and $g_{s}^{\ast}$ is the electron (hole) effective
Land\'{e} g-factor. Note that we lump heavy and light holes together
by the trace and adopt an average value of
$\mu_{B}^{\ast}g_{s}^{\ast}$ for the valence band.

\begin{figure}[t]
\includegraphics[width=8cm]{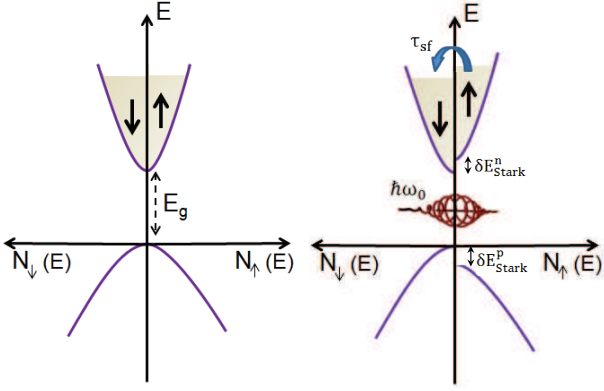}\caption{Illustration of the changes
in the majority and minority population due to the Stark shift ($\delta
E_{\mathrm{Stark}}^{\mathrm{n}}<\delta E_{\mathrm{Stark}}^{\mathrm{p}}$), in
the presence of non-resonant and intense circularly polarized laser field.}%
\label{stark}%
\end{figure}

As illustrated by Fig. \ref{stark}, the below-band gap light field
induces a Zeeman-like splitting, called the spin-selective optical
Stark shift \cite{shen}, which can be interpreted as an effective
magnetic field experienced by each band, $\delta
E_{\mathrm{Stark}}=-\mu_{B}^{\ast}g_{s}^{\ast}B_{\mathrm{IFE}}/2$.
The effective field, $B_{\mathrm{IFE}}$, is a reactive response and
therefore essentially instantaneous as long as $T_{p}>\hbar/E_{g}$.
The effective field eventually gives rise to a non-equilibrium
spin-polarization in the conduction band,
$\langle{\sigma}_{z}\rangle_{\mathrm{ne}}^{(n)}\propto N^{(n)}\delta
E_{\mathrm{Stark}}$,
where $N^{(n)  }$ is the density of states per unit volume at the
Fermi level of the conduction band (with analogous relations for the
holes) and $\langle{\ldots}\rangle$ denotes the expectation value on
intermediate time scales. This spin polarization is generated by a
re-population of the states (see Fig. \ref{stark}), on the scale
of the spin-flip scattering time $T_{p}\sim\tau_{sf}$, which is
expected to be in the ps range under high excitation conditions.

\subsection{Paramagnetic case}
In second-order time-dependent perturbation theory, the spin
susceptibility of a paramagnet is
defined as $\mathbf{K}_{\mathrm{IFE}}=\langle\boldsymbol{\sigma}%
\rangle_{\mathrm{ne}}/(\xi E_{0}^{2}/\omega_{0}^{2})$ with $\xi=e^{2}/\left(
mc^{2}\right)  $, which reads
\begin{align}
\mathbf{K}_{\mathrm{IFE}}  &  =-\frac{1}{L^{3}}\sum_{km}\frac{f(\epsilon
_{k})-f(\epsilon_{m})}{(\epsilon_{k}-\epsilon_{m})+i\eta}\langle
k|\boldsymbol{s}|m\rangle\times\label{final_K}\\
&  \left[  C_{\rho}\mathbf{{\hat{e}}}\cdot\mathbf{\hat{I}}\cdot\mathbf{{\hat
{e}}}^{\ast}\langle m|k\rangle+iC_{\sigma}(\mathbf{{\hat{e}}}\times
\mathbf{{\hat{e}}}^{\ast})\cdot\langle{m}|\boldsymbol{s}|{k}\rangle\right]
,\nonumber
\end{align}
where $L$ is the system size, $\eta$ is a positive infinitesimal,
$f(\epsilon)$ is the Fermi-Dirac distribution function in
equilibrium, and $\mathbf{\hat{I}}$ is the unit dyadic in Pauli spin
space. The interband couplings $C_{\sigma}$ and $C_{\rho}$ for
$n$-doped semiconductors are
\begin{align}
C_{\sigma}^{(n)}  &  =\frac{2P^{2}}{3m}\left(  \frac{-\hbar\omega_{0}}%
{E_{g}^{2}-\hbar^{2}\omega_{0}^{2}}+\frac{\hbar\omega_{0}}{(E_{g}+\Delta
)^{2}-\hbar^{2}\omega_{0}^{2}}\right)  ,\label{C_sigma}\\
C_{\rho}^{(n)}  &  =\frac{2P^{2}}{3m}\left(  \frac{2E_{g}}{E_{g}^{2}-\hbar
^{2}\omega_{0}^{2}}+\frac{E_{g}+\Delta}{(E_{g}+\Delta)^{2}-\hbar^{2}\omega
_{0}^{2}}\right)  , \label{C_rho}%
\end{align}
where $\Delta$ is the spin-orbit splitting energy and $P$ the
interband momentum matrix element. The coefficients of $C_{\sigma}$
and $C_{\rho}$ are identical to the spin- and charge-density
excitation coefficients in the theory of Raman scattering
\cite{pinczuk}. In our formulation, the incoming and outgoing
photons have identical polarization, Eq. (\ref{final_K}), which
means that there is no direct angular momentum transfer from the
light to the medium. Angular momentum of spin-flip processes is
hence supplied from the lattice via spin-orbit coupling during the
spin-flip relaxation process.

In a paramagnetic $n$-doped semiconductor, the Stark effective
field, or IFE field, is oriented along the light propagation
direction ${\hat{\mathbf{{q}}}}$, as
\begin{equation}
{{\mathbf{B}}}_{\mathrm{IFE}}^{(n)}=\frac{-2\xi E_{0}^{2}}{\mu_{B}^{\ast}%
g_{s}^{\ast}\omega_{0}^{2}}\frac{\mathbf{K}_{\mathrm{IFE}}}{N^{(n)}}%
=-\frac{2\lambda C_{\sigma}^{(n)}\xi E_{0}^{2}}{\mu_{B}^{\ast}g_{s}^{\ast
}\omega_{0}^{2}}{\hat{\mathbf{{q}}}}. \label{n-doped-field}%
\end{equation}
For $T_{p}\gtrsim\tau_{sf}$ this magnetic field leads to the spin accumulation
$\left\langle \boldsymbol{\sigma}\right\rangle _{\mathrm{ne}}^{(n)}%
=-K_{z,\mathrm{IFE}}\xi
E_{0}^{2}{\hat{\mathbf{q}}/}\omega_{0}^{2}=-2\lambda
N^{(n)}C_{\sigma}^{(n)}\xi
E_{0}^{2}{\hat{\mathbf{{q}}}/}\omega_{0}^{2}$.

In paramagnetic $p$-doped systems, the effective field experienced
by the hole bands are
\begin{equation}
\mathbf{{B}}_{\mathrm{IFE}}^{(p)}=-\frac{3\lambda C_{\sigma}^{(p)}\xi
E_{0}^{2}{\hat{\mathbf{{q}}}}}{\mu_{B}^{\ast}g_{s}^{\ast}\omega_{0}^{2}},
\label{p-doped-field}%
\end{equation}
where
$C_{\sigma,\rho}^{(p)}=C_{\sigma,\rho}^{(n)}(\Delta\rightarrow\infty)$
since the matrix elements\ with the spin-orbit split-off bands
vanish. In the $p$-doped case, the non-equilibrium spin polarization
on longer time scales is $\langle\boldsymbol{\sigma}\rangle_{\mathrm{ne}}^{(p)}=\lambda N^{(p)}%
C_{\sigma}^{(p)}\xi E_{0}^{2}{\hat{\mathbf{{q}}}}\mathbf{{/}}%
\omega_{0}^{2}$, where $N^{(p)}$ is the average density of states at
the Fermi level of hole bands.

According to Eqs. (\ref{n-doped-field}) and (\ref{p-doped-field}),
polarized light with frequency $\omega_{0}<E_{g}/\hbar$ induces a
magnetic field along ${\hat{\mathbf{q}}}$. Its sign is governed by
the light helicity $\lambda$, while its magnitude is proportional to
the light intensity
$E_{0}^{2}$ and vanishes with the spin-orbit coupling since $C_{\sigma}%
^{(n)}(\Delta=0)=0$. In $n$-doped systems and in the large detuning
limit $\omega_{0}\ll E_{g}/\hbar$, to leading order, it yields
$\langle\sigma^{z}\rangle_{\mathrm{ne}}\propto\Delta/\omega_{0}$ for
$\Delta<E_{g}$. This optical Stark shift-induced non-equilibrium
spin polarization can  be compared with the magnetization induced by
the circular currents in response to the rotating electric field of
the circularly polarized light \cite{kubo_plasma}. The latter scales
with frequency like $\propto\omega_{0}^{-3}$, thus should be small
at optical frequencies. The spin-transfer torques induced by the
circular currents might be significant, however \cite{Fateme}.

This perturbation theory is valid in the limit $\delta E_{\mathrm{Stark}%
},\varepsilon_{F}\ll E_{g}$, and $\hbar\omega_{0}<E_{g}$. For \textit{n}-GaAs
with $E_{g}=1.52$ $%
\operatorname{eV}%
$, $\Delta=341$ meV, $ g_{s}^{\ast}\simeq -0.44$,
$m^{\ast}\simeq0.067m$ and $2P^{2}/m\simeq20$ eV \cite{g-factor}, a
light
intensity of $10~\mathrm{%
\operatorname{GW}%
/}%
\operatorname{cm}%
^{2}$ at frequency $\hbar\omega_{0}=1.24~\mathrm{eV}~(\lambda_{0}=1~%
\operatorname{\mu m}%
)$ then generates an effective magnetic field of $9~\mathrm{T}$.
This estimate is more than three orders of magnitude larger than
what has been predicted in disordered metals involves intraband
transitions with the same laser intensities \cite{tatara,edelstein}.
The high-intensity laser power used in ultrafast opto-magnetic
experiments leads to heating and demagnetization of samples even at
below band gap frequencies due to multiple photon absorption, band
tails, disorder \textit{etc}. The optically induced spin
accumulation then can nucleate a persistent magnetization when the
samples cools after the pump pulse \cite{kimel1,kimel2}.

\subsection{Ferromagnetic case} Consider now hole-doped
\emph{ferromagnetic} semiconductors. As before, we assume small hole
densities $\varepsilon_{F}\ll E_{g},\Delta$ and thus limit the
discussion to the optical transitions at $\Gamma$. We investigate
the weak ferromagnetic regime in which $J\ll\Delta<E_{g}$, therefore
it is sufficient to calculate effective fields to the lowest order
of $J/\Delta$.

First, we assume an equilibrium magnetization direction
$\hat{\mathbf{{M}}}_{0}$ along the light propagation direction
$\hat{\mathbf{{q}}}$.
The average effective field experienced by the valence bands is
\begin{align}
\mathbf{{B}}_{\mathrm{light}}^{(p)}  &  =\mathbf{{B}}_{\mathrm{IFE}}%
^{(p)}+\mathbf{{B}}_{\mathrm{ICME}}^{(p)}\nonumber\\
&  \simeq-\left(  3\lambda C_{\sigma}^{(p)}+\frac{2J}{5\Delta}C_{\rho}%
^{(p)}\right)  \frac{\xi E_{0}^{2}{\hat{\mathbf{{q}}}}}{\mu_{B}^{\ast}%
g_{s}^{\ast}\omega_{0}^{2}}. \label{lh_field}%
\end{align}
The first term, the helicity ($\lambda$) dependent term, is the IFE
field. The helicity independent term, the second term, corresponds
to the ICME field and enhances or suppresses the magnetization even
for a linearly polarized beam. The ICME is an odd function of the
exchange coupling and, in the small magnetization limit, is linearly
proportional to the exchange coupling. This is in contrast to the
IFE, which is even in the exchange energy and odd in the helicity.
In the large detuning limit, the ICME scales like $\omega_{0}^{-2}$,
while the helicity dependent IFE is $\propto\omega _{0}^{-1}$. In
this case $\hat{\mathbf{{q}}}\Vert\hat{\mathbf{{M}}}_{0}$, and then
the effective field does not trigger magnetization precession
dynamics. Linear reversal through zero magnetization can occur if
the light-induced effective field is sufficiently larger than
coercive field, which is dramatically reduced down to a few teslas
near the Curie temperature \cite{nowak}. Magnetization reversal by
precession might be possible in principle, but would require much
longer light pulses for
$\hat{\mathbf{{q}}}\Vert\hat{\mathbf{{M}}}_{0}$.

\begin{figure}[t]
\includegraphics[width=5.5cm]{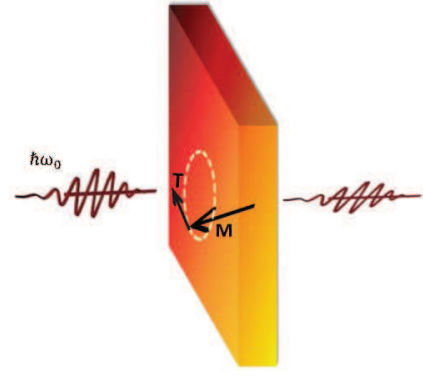}\caption{Both linearly and circularly
polarized light exert a torque $\mathbf{{T}}$, on the equilibrium
magnetization $\mathbf{{M}}$, and may trigger magnetization dynamics.}%
\label{stark2}%
\end{figure}

Second, we consider a magnetization direction perpendicular to the light
propagation direction as $\hat{\mathbf{{M}}}_{0}=\hat{\mathbf{{x}}}\cos
\phi+\hat{\mathbf{{y}}}\sin\phi$, where $\phi$ is the azimuthal angle. The
average light-induced effective field induced by circularly polarized light on
the holes in the valence bands is
\begin{align}
\mathbf{{B}}_{\mathrm{light}}^{(p)}  &  =\mathbf{{B}}_{\mathrm{IFE}}%
^{(p)}+\mathbf{{B}}_{\mathrm{ICME}}^{(p)}\nonumber\\
&  \simeq-\left(  3\lambda C_{\sigma}^{(p)}\hat{\mathbf{{q}}}-\frac{J}%
{5\Delta}C_{\rho}^{(p)}\hat{\mathbf{{M}}}_{0}\right)  \frac{\xi E_{0}^{2}}%
{\mu_{B}^{\ast}g_{s}^{\ast}\omega_{0}^{2}}. \label{Mxy-cir}%
\end{align}
The first term, the IFE field, is along the light wave vector and
changes sign with light helicity. The IFE field acts on the holes
that relax very fast to generate a spin accumulation
$\langle\bm{\sigma}\rangle^{(p)}_{\mathrm{ne}}$, which by the
exchange interactions exerts a strong torque on the local
magnetization
$\mathbf{{T}}=J{\hat{\mathbf{{M}}}}_{0}\times\langle\bm{\sigma}\rangle^{(p)}_{\mathrm{ne}}
\propto
{\hat{\mathbf{{M}}}}_{0}\times\mathbf{{B}}^{(p)}_{\mathrm{light}}$
\cite{matos}. The other term corresponds to the ICME and is strictly
longitudinal, which enhances or suppresses magnetization, but does
not trigger magnetization precession \cite{kawasaki}. With
$J/\Delta\sim0.1$ we estimate
$B_{\mathrm{ICME}}^{(p)}\sim10^{-4}%
\operatorname{eV}%
/\left(  \mu_{B}^{\ast}g_{s}^{\ast}\right)$ for a light intensity of
$10~\mathrm{%
\operatorname{GW}%
/}%
\operatorname{cm}%
^{2},$ which is large considering that
$\mu_{B}^{\ast}g_{s}^{\ast}\sim 10^{-4}-10^{-5}$ eV/T. \\Also the
effective magnetic field induced by linearly polarized light, in
perpendicular configuration $\mathbf{{q}}
\bot\hat{\mathbf{{M}}}_{0}$, is given by
\begin{align}
\mathbf{{B}}_{\mathrm{light}}^{(p)}  &  =\mathbf{{B}}_{\mathrm{ICME}}%
^{(p)}+\mathbf{{B}}_{\mathrm{\bot}}^{(p)}\nonumber\\
&  \simeq\frac{2J}{5\Delta}C_{\rho}^{(p)}\left[  \hat{\mathbf{{M}}}_{0}\left(
3\cos^{2}(\phi-\alpha)-1\right)  \right. \nonumber\\
&  \left.  +\frac{3}{2}\hat{\mathbf{{q}}}\times\hat{\mathbf{{M}}}_{0}%
\sin2(\phi-\alpha)\right]  \frac{\xi E_{0}^{2}}{\mu_{B}^{\ast}g_{s}^{\ast
}\omega_{0}^{2}}. \label{Mxy-lin}%
\end{align}
This field has two components, the conventional ICME parallel to $\hat{\mathbf{{M}}%
}_{0}$, and a term along $\hat{\mathbf{{q}}}\times\hat{\mathbf{{M}}}_{0}%
$, which exerts a torque on the local magnetization in the
$z$-direction. Eqs. (\ref{Mxy-cir}) and (\ref{Mxy-lin}) show that in
the perpendicular configuration both linearly and circularly
polarized light induce effective fields that exert torques on the
equilibrium magnetization and induce precessional dynamics. Note,
however, that in our model, unpolarized light or just a heat pulse
does not generate effective magnetic fields. Linearly polarized
light does not carry net angular momentum, but nevertheless induces
spin precession by inducing angular momentum transfer between
lattice and exchange field, thereby rotating its plane of
polarization (Faraday effect), see Fig. \ref{stark2}. Circularly
polarized photons can directly transfer angular momentum from the
light to the spin of electrons only when absorbed. At typical laser
intensities, the amount of available angular momentum is by far not
enough to reverse the magnetization. We thus demonstrate that in
optomagnetism the lattice and exchange fields act as sources and
sinks of angular momentum via the spin-orbit and exchange couplings
\cite{koopman}. We present here a microscopic theory of
light-induced magnetic fields. In order to compare with experiments,
the magnetization dynamics under effective magnetic field and heat
pulses will have to be computed. A realistic micromagnetic
simulation in the presence of such an effective field has been
carried out in Refs. [5] and [27]. A repetition of these
calculations for III-V magnetic semiconductors is far beyond the
scope of this paper, however.

\section{SUMMARY AND CONCLUSIONS}
In summary, we studied the magnetic response to in- tense and
non-absorptive, linearly and circularly polarized lights in para-
and ferromagnetic III-V semiconductors. The strong spin-orbit
coupling plays a vital role to supply the required angular momentum.
As a result, the light-induced field strength in GaMnAs is huge, up
to several teslas, which is sufficient to reverse magnetization by
either linearly or precessional paths. We found that the
spin-selective optical Stark effect in ferromagnets induces
effective magnetic fields in different directions depending on the
light orientation and the magnetization direction.

\section{Acknowledgments}
This work was supported by EU-ICT-7 contract No. 257159
\textquotedblleft MACALO\textquotedblright, the FOM Foundation, the
ICC-IMR and DFG Priority Programme 1538 \textquotedblleft
Spin-Caloric Transport\textquotedblright. A. Q. would like to thank
A. G. Moghaddam for useful discussions.

\end{document}